\documentclass[11pt,english]{iopart} 

\usepackage[usenames,dvipsnames]{color}
\usepackage[normalem]{ulem}
\usepackage{graphicx}

\expandafter\let\csname equation*\endcsname\relax
\expandafter\let\csname endequation*\endcsname\relax

\usepackage{amsmath}
\usepackage{amssymb,amsthm}
\usepackage{amsfonts}
\usepackage{color}
\usepackage{graphics}
\usepackage{epsfig}
\usepackage{bbm}
\usepackage{pstricks}
\usepackage{subfigure}
\usepackage{cite}

\newcommand{\be}{\begin{equation}}
\newcommand{\ee}{\end{equation}}
\newcommand{\eea}{\end{eqnarray}}
\newcommand{\bea}{\begin{eqnarray}}

\newcommand{\eins}{1\!\!1}
\renewcommand{\qed}{\ensuremath{\hfill \Box}}

\newcommand{\ketbra}[1]{\ensuremath{| #1 \rangle \!\langle #1 |}}
\newcommand{\ketbras}[1]{\ensuremath{| #1 \rangle \!\langle \cdot |}}

\newcommand{\ket}[1]{\ensuremath{|#1\rangle}}

\newcommand{\braket}[2]{\ensuremath{\langle #1|#2\rangle}}

\newcommand{\kommentar}[1]{}
\newcommand{\trace}{{\rm tr}}

\newcommand{\vr}{\ensuremath{\varrho}}

\newcommand{\forget}[1]{}

\begin{document}


\title{Analytical characterization of the genuine 
multiparticle negativity}

\author{Martin Hofmann, Tobias Moroder, and Otfried G\"uhne}
\address{Naturwissenschaftlich-Technische Fakult\"at,
Universit\"at Siegen,
Walter-Flex-Str.~3,
57068 Siegen, Germany}
\ead{hofmann@physik.uni-siegen.de}

\date{\today}


\begin{abstract}
The genuine multiparticle negativity is a measure of genuine multiparticle
entanglement which can be numerically calculated. 
We present several results how this entanglement measure
can be characterized in an analytical way. First, we show that with
an appropriate normalization this measure can be seen as coming from
a mixed convex roof construction. Based on this, we determine its 
value for $n$-qubit GHZ-diagonal states and four-qubit cluster-diagonal 
states.
\end{abstract}




\section{Introduction}

Entanglement is believed to be a very useful resource in quantum 
information processing. It is involved in some quantum key distribution 
protocols, quantum metrology, quantum phase transitions and many 
other physical applications and phenomena. Therefore it is one of the 
main tasks to detect and quantify entanglement, especially in the 
multiparticle setting. One tool to quantify genuine multiparticle 
entanglement is the genuine multiparticle negativity (GMN), introduced 
in Ref.~\cite{Jungnitsch11}. 

The GMN is easily computable via semidefinite programming and has been 
useful to study a large variety of questions. It was used to identify 
genuine multiparticle entanglement in the state of the generated photons 
of the triple Compton effect \cite{compton}, the high-energy process 
in which a photon splits into three after colliding with a free electron. 
Further, the GMN allowed to quantitatively study the scaling and spatial 
distribution of genuine multiparticle entanglement in the reduced states 
of a one-dimensional spin model at a quantum phase transition \cite{GMEscaling}. 
In Ref.~\cite{GMEindisssys} it helped along with other methods to track the 
dynamics of the entanglement structure of a multiparticle open quantum system 
from genuine multiparticle entanglement to full separability and in 
Ref.~\cite{decoh} it was used to study the robustness of different types
of entangled states against decoherence. In Ref.~\cite{GMEinnetwork},
the GMN quantified how quantum reservoir engineering can create entangled states 
in cascaded quantum-optical networks driven by dissipative processes. Finally, 
it has been shown how the GMN can be measured experimentally in a 
device-independent way \cite{PRL.111.030501}.
All these
applications are based on the fact that the GMN can directly be computed, but this
may also lead to the impression that the GMN is mainly a numerical tool and not
accessible to an analytical treatment.

In this paper we present an analytical study of the GMN. First, we show that 
a renormalized version of the GMN can be expressed as mixed convex roof of 
the minimum of bipartite  negativities \cite{VidalWerner}. These mixed 
convex roofs were already studied in the context of entanglement quantification 
in the bipartite setting in Refs.~\cite{mixedroof}. In our case and contrary 
to the usual pure state convex roof constructions the renormalized GMN can 
be efficiently computed using semidefinite programming.  Second, we derive 
analytic expressions for the GMN two different state families. These are 
the GHZ-diagonal $n$-qubit and the cluster-diagonal four-qubit states. 
These analytic formulas for the GMN in terms of the fidelities of the 
GHZ and cluster states also provide lower bounds on the 
genuine multiparticle entanglement of general mixed quantum states.

This paper is organized as follows. In Section 2 we review basic 
notions such as genuine multiparticle entanglement, PPT mixtures 
and the genuine multiparticle negativity. In Section 3 we introduce 
the renormalized GMN and show that it can be expressed as a mixed 
state convex roof. We then compare the original GMN and the 
renormalized GMN and provide the naturally arising upper and 
lower bounds to the latter. In Section 4 we derive an analytic 
formula for the original and renormalized GMN for $n$-qubit 
GHZ-diagonal states and compare our results to the GME-concurrence 
\cite{GMECX, GME-concurrence}. 
We also show that an exact expression can also be obtained for 
cluster-diagonal four-qubit states, where only lower bounds are 
known for the GME-concurrence 
\cite{GME-concurrence,PhysRevA.86.022319,PhysRevA.83.040301}. We conclude 
our work with a brief discussion of our results and a short outlook 
on possible future directions.

\section{Basic definitions}
Before we can start to present our results, we recall the basic 
notions of genuine multiparticle entanglement, PPT mixtures and 
the multiparticle negativity.

\subsection{Genuine multiparticle entanglement}
We consider for simplicity three parties, Alice (A), Bob (B) 
and Charlie (C), the generalization to more parties is 
straightforward. A three-particle state is fully separable 
and contains no entanglement, if it can be written as a 
mixture of product states
\begin{equation}
\varrho^{\mathrm{fullsep}} = \sum_k p_k \ketbra{\psi^k_A} \otimes \ketbra{\phi^k_{B}} \otimes \ketbra{\phi^k_{C}},
\end{equation}
where the $p_k$ form a probability distribution. For states which are
not of this form there are different types of entanglement. It may happen that the particles $A$ and $B$ are 
entangled, whereas particle $C$ is separable. Such a state is said 
to be separable with respect to the bipartition $C$ versus $AB$ 
($C\vert AB$) and a mixture of product states of the 
joint system $AB$ and is the single system $C$, 
\begin{equation}
\varrho^{\mathrm{sep}}_{C\vert AB} = \sum_k p_k \ketbra{\psi^k_{AB}}\otimes\ketbra{\phi^k_{C}}.
\end{equation}
Similarly one defines states which are separable with respect to 
the other possible splittings $B \vert AC$ and $A\vert BC$. The 
convex hull of these three sets defines the set of biseparable 
states,
\begin{equation}
\varrho^{\mathrm{bisep}} = p_{A}\varrho^{\mathrm{sep}}_{A\vert BC} + p_{B}\varrho^{\mathrm{sep}}_{B\vert AC} + p_{C}\varrho^{\mathrm{sep}}_{C\vert AB}.
\label{eqn:bisep}
\end{equation}
States outside this set are called genuine multiparticle 
entangled and are the states of interest for many experimental applications.

\begin{figure}[t]
\begin{center}
\subfigure[]{\includegraphics[width=0.40\textwidth]{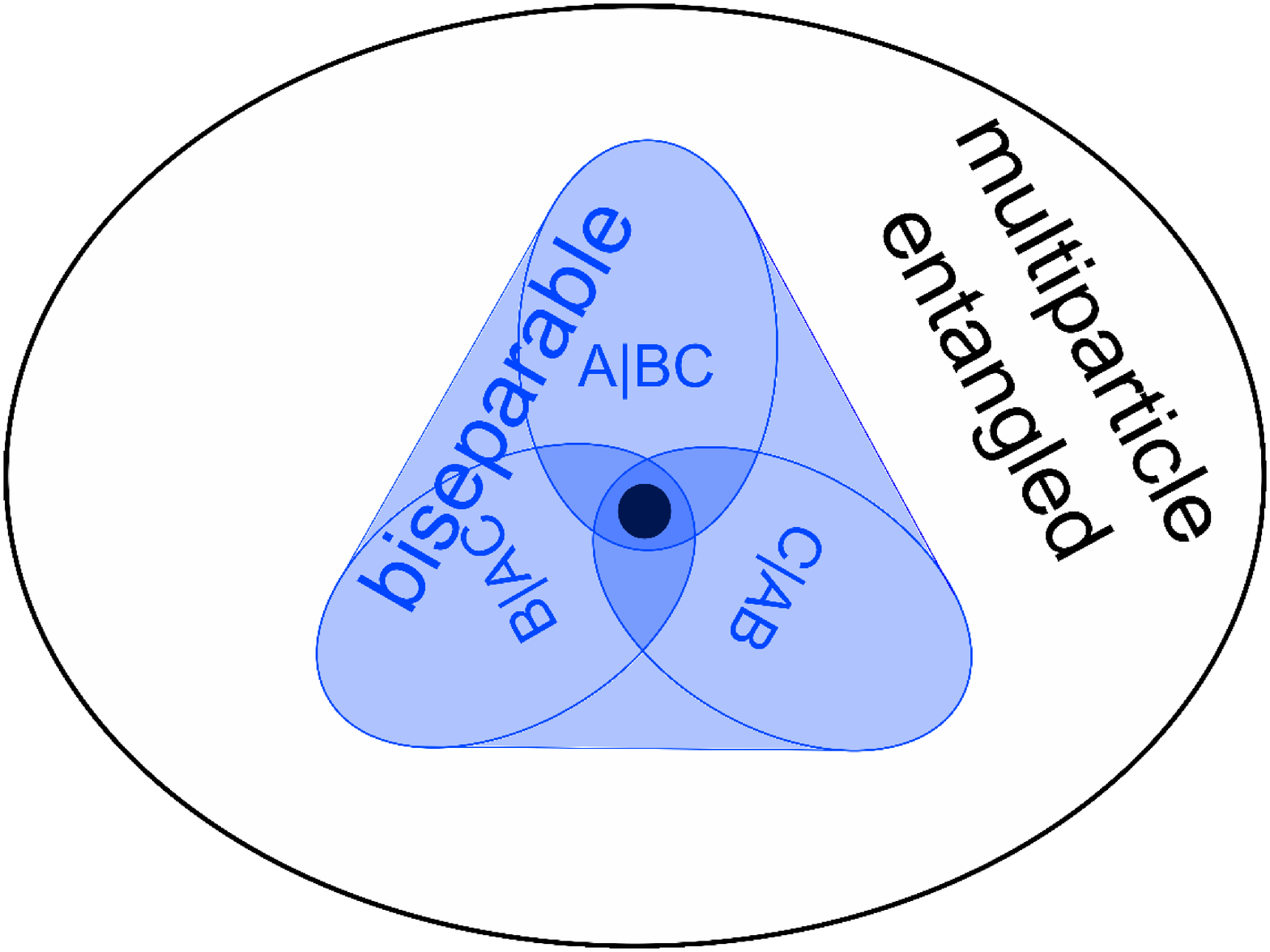}} 
\hspace*{0.5cm}
\subfigure[]{\includegraphics[width=0.40\textwidth]{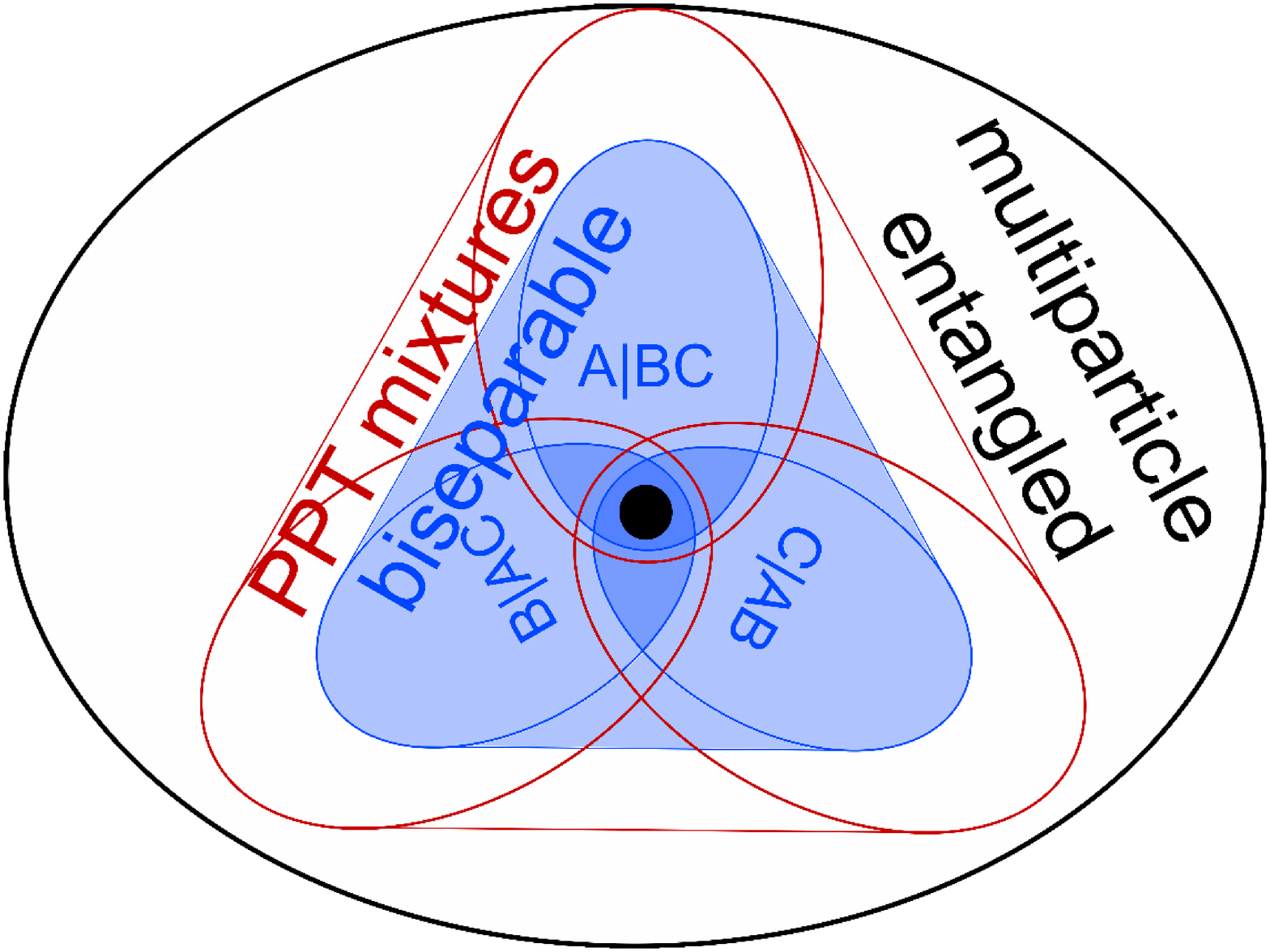}}
\end{center}
\caption{(a) The nested structure of the different convex sets of 
separable states within the set of all states. The innermost set (dark dot) represents the fully separable states. The three sets of states which are separable (with respect to a partition $A\vert BC$, $B\vert AC$ or $C\vert AB$) are supersets of the set of fully separable states, which is contained in but not equal to the intersection of the three. The biseparable states form the convex hull of the separable states with respect to the different partitions. All states outside the set of biseparable states are genuine multiparticle entangled. (b) The PPT mixtures are the convex hull of states, which are PPT with respect to a single partition. As all states which are separable with respect to a certain partition are also PPT, the PPT mixtures form a superset of the biseparable states. States which are not PPT mixtures are genuine multiparticle entangled.}
\label{fig:nestedsep}
\end{figure}

In the three-particle case one ends up with a nested structure 
of different kinds of separable states, all of which are contained 
within the biseparable states (see Fig. \ref{fig:nestedsep}(a)).

\subsection{PPT mixtures and the genuine multiparticle negativity}
At present, there is no general framework to prove or disprove 
the existence of a biseparable decomposition for arbitrary mixed 
states. In Ref.~\cite{Jungnitsch11} this problem was studied by 
introducing a relaxation. Instead of trying to characterize the 
set of biseparable states as given in Eq.~(\ref{eqn:bisep}) 
one characterizes the superset of so-called PPT mixtures. 

The idea  builds on the fact that the separable states 
are a subset of the  states with a positive partial transpose 
(PPT states) \cite{pptcriterion}. Recall that a bipartite state 
\begin{equation}
\varrho_{A\vert B} = \sum_{ij,kl} \varrho_{ij,kl} \vert i\rangle\langle j\vert_A \otimes \vert k\rangle\langle l\vert_B 
\label{eqn:bipartitestate}
\end{equation}
is called PPT, if its partial transposition with respect to the 
subsystem $A$,
\begin{equation}
\varrho_{A\vert B}^{T_A} = \sum_{ij,kl} \varrho_{ij,kl} \vert j\rangle\langle i\vert_A \otimes \vert k\rangle\langle l\vert_B 
\label{eqn:bipartitestateTA}
\end{equation}
has no negative eigenvalues.

In the three-particle case [see Fig.~\ref{fig:nestedsep}(b)]
PPT mixtures are states, which can be written as
\begin{equation}
\varrho^{\rm PPTmix} = p_{A}\varrho^{\mathrm{PPT}}_{A\vert BC} 
+ p_{B}\varrho^{\mathrm{PPT}}_{B\vert AC} + p_{C}\varrho^{\mathrm{PPT}}_{C\vert AB}.
\end{equation}
A general PPT mixture on more than three parties will have $2^{n-1}-1$ terms, one for each partition $ m\vert \bar m$ of the system into two parts. Such a bipartition is a splitting of the system into a part $m$ and 
its complement $\bar m$. Note, however, that $m\vert \bar m$ and 
$\bar m \vert m$ label the same bipartition.

The main advantage of this approach is that for any given multiparticle state 
$\varrho$ one can directly check whether it is a PPT mixture or 
not. For that, it was shown that the non-existence of such a 
decomposition is equivalent to the existence of a fully decomposable 
witness $\mathcal W$ detecting the state $tr(\mathcal W\varrho)<0$. 
Such a witness is an operator $\mathcal W$, which can be written 
for all possible bipartitions $m\vert \bar{m}$ as 
$\mathcal W = P_m+Q_m^{T_m}$, with positive operators $P_m$ and 
$Q_m$. Finding such a witness can be cast into a so-called semidefinite
program (SDP, see also below), which can be solved efficiently with 
standard numerical routines \cite{SDP, pptmixerprogram}.

Building upon this idea, Ref.~\cite{Jungnitsch11} introduced a 
computable entanglement monotone called the genuine multiparticle negativity (GMN). The basic idea
is to take a fully decomposable witness as above, and use the violation
of it as a quantifier of entanglement.  More precisely, one defines
the GMN $\tilde{N}_g(\varrho)$ via the 
optimization problem:
\begin{eqnarray}
\label{eqn:sdpgennegold}
\tilde{N}_g(\varrho) &= 
-\min \trace\left( \varrho \mathcal W \right)\\
\nonumber &          \mbox{ subject to: }   \mathcal W = P_m +Q_m^{T_m},\\
\nonumber & \phantom{\mbox{ subject to: }} 0\le P_m \le\eins,\\
\nonumber & \phantom{\mbox{ subject to: }} 0\le Q_m \le\eins\ \mbox{ for all partitions } m\vert \bar{m}.
\end{eqnarray}
For the three-particle case the witness operator has to be decomposable into 
$\mathcal W=P_A+Q_A^{T_A}$, $\mathcal W=P_B+Q_B^{T_B}$ and $\mathcal W=P_C+Q_C^{T_C}$ 
with $0 \le P_m,Q_m \le \eins$. Since this measure is defined as an optimization 
over a set of  witnesses it is closely related to the approach to quantify
entanglement based on entanglement witnesses as in Ref.~\cite{PRA.72.022310}.
In contrast to the existing approaches, however, it can be directly computed
using SDP \cite{pptmixerprogram}, since the optimization problem in Eq.~(\ref{eqn:sdpgennegold}) is directly an optimization problem of 
this class.
We finish this section by recalling the main properties of the GMN as proved in Refs.~\cite{Jungnitsch11, PRA.84.032310}:

\noindent
{\bf Lemma 1. }
{\it
The measure $\tilde N_g$ as defined by Eq. (\ref{eqn:sdpgennegold}) satisfies:
\begin{enumerate}
\item $\tilde{N}_g$ vanishes on all biseparable states 
$\varrho^{\mathrm{bisep}}$, i.e. $\tilde{N}_g(\varrho^{\mathrm{bisep}} ) = 0$. Further, if $\vr$ is no PPT mixture, then $\tilde{N}_g(\vr) > 0$.

\item $\tilde{N}_g$ is non-increasing under full LOCC operations (no joint operations on more than one part are allowed), i.e. $\tilde{N}_g(\Lambda_{\mathrm{LOCC}}(\varrho))\le \tilde{N}_g(\varrho)$.

\item $\tilde{N}_g$ is invariant under local basis changes $U_{\mathrm{loc}}$, i.e. $\tilde{N}_g(U_{\mathrm{loc}}\varrho U_{\mathrm{loc}^\dagger}) = \tilde{N}_g(\varrho)$.

\item $\tilde{N}_g$ is convex, i.e. $\tilde{N}_g(\vr) \le \sum_i p_i \tilde{N}_g(\varrho_i)$ holds for all convex decompositions $\vr=\sum_i p_i \varrho_i$.

\item $\tilde{N}_g$ is bounded by $\tilde{N}_g(\varrho) \le \frac 1 2 (d_{\mathrm{min}}-1)$, where $d_{\mathrm{min}}$ is the lowest dimension of any particle in the system \cite{PRA.84.032310}.
\item If the system consists of two parties only, then $\tilde{N}_g$ equals the bipartite negativity \cite{VidalWerner}.
\end{enumerate}
}

\section{The genuine multiparticle negativity as a convex roof measure}
In this section we introduce a renormalized version the GMN 
by changing the normalization of the witness operator. Our main 
motivation is the following. So far, we have a good understanding 
of the GMN in terms of witnesses. In state space however, there is 
no satisfying interpretation. By slightly altering the definition 
the GMN has a direct simple interpretation in the witness {\it and} 
in the state space picture. 

\subsection{Modifying the definition of the genuine multiparticle negativity}
For any state $\varrho$ the renormalized GMN 
$N_g(\varrho)$ is given by
\begin{eqnarray}
\label{eqn:sdpgenneg}
{N}_g(\varrho) &= 
-\inf \trace\left( \varrho \mathcal W \right)\\
\nonumber &          \mbox{ subject to: }   \mathcal W = P_m +Q_m^{T_m},\\
\nonumber & \phantom{\mbox{ subject to: }} 0\le P_m \\
\nonumber & \phantom{\mbox{ subject to: }} 0\le Q_m \le\eins\ \mbox{ for all partitions } m\vert \bar{m}.
\end{eqnarray}
Compared to the original definition in Eq.~(\ref{eqn:sdpgennegold}), 
the only difference is a relaxation in the constraints on the positive 
operators $P_m$, which is not bounded by $\eins$ anymore. Note 
that this definition was already used in Ref.~\cite{PRL.111.030501} 
to quantify genuine multiparticle entanglement in a device independent manner.

The interesting point is that the renormalized GMN has an interpretation
in state space as coming from an optimization over decompositions
of the density matrix $\varrho$. This is known as the mixed convex roof 
construction \cite{mixedroof}, and many entanglement measures are defined via such an 
optimization. In the present case, one deals with a so-called mixed 
convex roof, and this can be derived from the 
the dual problem \cite{SDP} to the semidefinite 
problem in Eq.~(\ref{eqn:sdpgenneg}). We have:

\noindent
{\bf Theorem 2. }{\it 
Let $N_m$ be the bipartite negativity given 
by $N_m(\varrho) = \sum_i \vert \lambda_i^-(\varrho^{T_m}) \vert$, where $\lambda_i^-(\varrho^{T_m})$ are the negative eigenvalues of $\varrho^{T_m}$. Then the genuine multiparticle 
negativity equals a mixed convex roof of bipartite negativities. 
That is
\begin{equation}
N_g(\varrho) = \min_{\varrho=\sum_m p_m\varrho_m} \sum_m p_m N_m(\varrho_m),
\label{eqn:genuinenegativity}
\end{equation}
where the summation runs over all inequivalent partitions
$m\vert \bar m$ of the system and the minimization is performed over all mixed
state decompositions of the state $\varrho=\sum_m p_m\varrho_m$. 
}

The proof of this Theorem can be found in \ref{sec:AA}.

Note that the optimization in Eq.~(\ref{eqn:genuinenegativity}) can also be written in a different
way: If one defines for an arbitrary multiparticle quantum state
the quantitity $\mu (\vr) = \min_{m} N_m (\varrho)$
as the bipartite negativity, minimized over all bipartitions, then
the multiparticle negativity can be written as 
\begin{equation}
N_g(\varrho) = \min_{\varrho=\sum_k p_k\varrho_k} \sum_k p_k \mu(\varrho_k),
\end{equation}
where now the minimization is over all decompositions $\varrho=\sum_k p_k\varrho_k$ into mixed states and $k$ does not label the bipartitions anymore. In this way, the connection to the usual convex roof construction (see Ref.~\cite{pureconvroof} and also Eq.~(\ref{gme-conv}) below) becomes more transparent. In general however mixed and pure state convex roofs are extremely difficult to compute. In this respect it is important to highlight that the renormalized GMN can be computed using SDP.

\subsection{Comparison with the original definition of the GMN}

First, we can state that all the properties of the GMN also hold
for the renormalized definition and one additional property is new.

\noindent
{\bf Lemma 3.}
{\it 
For the multiparticle negativity $N_g$ as defined in 
Eq.~(\ref{eqn:sdpgenneg}) all the properties (i) to (vi) 
listed in Lemma 1 hold. Additionally, it has 
the following property:
\begin{enumerate}
\setcounter{enumi}{6}
\item If $\ket{\psi}$ is a pure state, then 
\begin{equation}
\label{eqn:gennegpure}
N_g(\ket{\psi}) = \min_m N_m(\ket{\psi}),
\end{equation}
 where the minimization is performed over all bipartite 
 splittings $m\vert \bar{m}$ of the system.
\end{enumerate}
}

{\it Proof.}
The first properties from Lemma 3 can be proved 
directly as in Lemma 1 by modification of the respective proofs in Ref.~\cite{Jungnitsch11}. Concerning statement (vii), note that
for a pure state $\vr=\ketbra{\psi}$ there is only a single (and trivial)
decomposition, namely $\vr= 1 \cdot \ketbra{\psi}.$
\qed

Note that due to property (ii) of Lemmata 1 and 3 
both versions of the GMN 
are entanglement monotones. Naturally, the question 
arises how the renormalized GMN 
performs in detecting genuine multiparticle entangled states. 
If one is interested in the quantification of genuine multiparticle entanglement with the help of an entanglement monotone  
then either of the two monotones can be used as they are 
nonzero on the same set of states.\footnote{One 
can modify the original MATLAB implementation of the PPT mixer 
\cite{pptmixerprogram} to get an implementation of the renormalized GMN, by simply commenting line 90 in 
the ``entmon.m'' file.} One directly has:

\noindent
{\bf Corollary 4.}
{\it
For all $\varrho$, $\tilde{N}_g(\varrho)\le N_g(\varrho)$ and $\tilde{N}_g(\varrho)=0$ $\Leftrightarrow$ $N_g(\varrho)=0$.
}

Another feature of the renormalized GMN are the natural upper {\it and} 
lower bounds, arising from Eq.~(\ref{eqn:sdpgenneg}) and Eq.~(\ref{eqn:genuinenegativity}) in Theorem 2. First, as for the original GMN, every witness 
$\mathcal W$, which satisfies the constraints in Eq.~(\ref{eqn:sdpgenneg}) provides a lower bound on the renormalized genuine multiparticle 
negativity
\begin{equation}
- tr\left( \mathcal W\varrho \right) \le N_g(\varrho).
\label{eqn:lowerbound}
\end{equation}
Second, every mixed state decomposition of a state 
$\varrho = \sum_m p_m \varrho_m$, $p_m\ge 0$, $\sum_m p_m = 1$ 
provides an upper bound on the renormalized GMN
\begin{equation}
\sum_m p_m N_m(\varrho_m) \ge N_g(\varrho).
\label{eqn:upperbound}
\end{equation}
This property makes the renormalized GMN easier to compute analytically. Note that the upper bounds of the renormalized GMN also provide upper bounds for $\tilde{N}_g$ since $\tilde{N}_g\le N_g$.

Finally, note that for pure states the renormalized GMN can directly
be computed with the help of Lemma 3. Sometimes it coincides with the
original GMN for pure states, and sometimes not.  An example is  the 
three-qubit GHZ state $\ket{GHZ} = 1/\sqrt{2}(\ket{000}+\ket{111}$, 
where $N_g(\ket{GHZ})=\tilde{N}_g(\ket{GHZ})=1/2$. On the other hand, 
for the three-qubit W state $\ket{W} = 1/\sqrt{3}(\ket{001}+\ket{010}+\ket{100})$, $N_g(\ket{W}) = \sqrt{2}/3 \approx 0.47$ and $\tilde{N}_g(\ket{W})\approx 0.43$.

\section{Analytic computation of the genuine multiparticle negativity}
In this section we use our previous results to provide analytic formulae 
of the GMN for two important families of multi-qubit states. These 
are the $n$-qubit GHZ-diagonal and four-qubit cluster-diagonal states. 
The idea in both cases is to construct for each family of states a 
family of witnesses lower bounding the GMN 
and a family of decompositions, which results in upper bounds. Since the bounds coincide and hold true for the original and the renormalized GMN they provide closed formulas for both monotones.

\subsection{Graph states}
\begin{figure}[t]
\begin{center}
\includegraphics[width=0.4\textwidth]{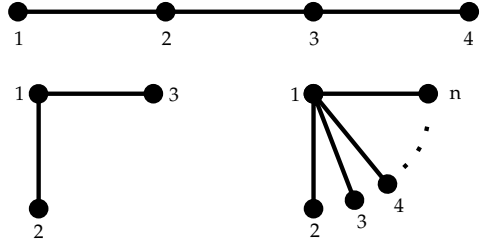}
\end{center}
\caption{Several examples for graphs.
Top: the linear cluster graph with four vertices. 
Bottom left: the star graph with with three vertices. 
This one coincides with the linear cluster graph having 
three vertices. 
Bottom left: the general $n$-vertex star graph.}
\label{fig:graph}
\end{figure}

Both state families are connected to so called graph states, which are relevant in many applications in quantum information processing \cite{Graphstatereview}. 
One defines them as follows: consider a graph $G$ which is a set of $n$ vertices with edges connecting them
(see Fig.~\ref{fig:graph}), where each vertex corresponds to a qubit. 
For each vertex $i$ we define its neighbourhood $\mathfrak N(i)$ by 
the set of all vertices, which are connected to $i$. 
Denote by $X^{(i)}$, $Y^{(i)}$ and $Z^{(i)}$ the Pauli matrices 
$\sigma_x$, $\sigma_y$ and $\sigma_z$ on the $i$-th particle with 
the identity on all the other. Then we can define for each vertex 
a so-called stabilizing operator
\begin{equation}
g_i := X^{(i)} \bigotimes_{j \in \mathfrak N(i)} Z^{(j)}.
\label{eqn:graphstabilizer}
\end{equation}
The graph state $\ket{G}$ is the unique $n$-qubit eigenstate 
to eigenvalue $+1$ to all $g_i$.
\begin{equation}
g_i\ket{G} = \ket{G}, \mbox{for all } i=1,2,\dots,n.
\label{eqn:graphstate}
\end{equation}
One can extend this framework by considering all common eigenstates 
of the stabilizing operators. We label those $2^n$ different states 
by their eigenvalues of $\pm 1$ on the stabilizing operators $g_k$, 
such that $g_i \ket{ a_1 a_2 \dots a_n} = a_i \ket{a_1 a_2 \dots a_n}$ 
with $a_i = \pm$. Note that these states are all orthogonal 
$\braket{a_1 \dots a_n}{b_1 \dots b_n} = \prod_{i=1}^N \delta_{a_i b_i}$ 
and thus form a basis in the $n$-qubit Hilbert space, the so-called
graph state basis. Mixed states, which are diagonal in the graph state 
basis are determined by their fidelities 
$F_{a_1 a_2 \dots a_n} = \langle a_1 a_2 \dots a_n \vert \varrho \vert a_1 a_2 \dots a_n \rangle$
\begin{equation}
\varrho = \sum_{a_1,a_2,\dots,a_n} F_{a_1 a_2 \dots a_n} \ketbra{a_1 a_2 \dots a_n}.
\label{eqn:graphdiag}
\end{equation}
They are called graph-diagonal and have the property that they are 
invariant under the group generated by the stabilizing operators, 
i.e. $g_i \varrho g_i^\dagger = \varrho$ for all $i$.

Note that an arbitrary state $\vr$ can be transformed into a graph-diagonal state by the symmetrization operation
\begin{equation}
\vr_{graph-diag} = \frac{1}{2^n} \sum_{g\in G} g\vr g^\dagger,
\label{eqn:symmetrization}
\end{equation}
were the summation runs over all group elements of the group generated by the $g_i$. Since all $g\in G$ consist of local Pauli operators only this symmetrization does not increase entanglement and hence $N_g(\vr_{graph-diag}) \le N_g(\vr)$.

\subsection{$n$-qubit GHZ-diagonal states}

In this paragraph we consider generalized Greenberger-Horne-Zeilinger
(GHZ) states. That are states, which are diagonal in the $n$-qubit 
GHZ-basis consisting of $2^n$ states $\ket{\psi_i} = 1 / \sqrt{2} \left( \ket{x_1x_2\dots x_n}\pm\ket{\bar{x}_1\bar{x}_2\dots \bar{x}_n} \right)$, where $x_j, \bar{x}_j\in \left\{ 0,1 \right\}$ and $x_j\not = \bar{x}_j$.
In the three-qubit case this 
basis  consists of the states 
$1/\sqrt{2}\left( \vert 000 \rangle \pm \vert 111 \rangle \right)$, 
$1/\sqrt{2}\left( \vert 001 \rangle \pm \vert 110 \rangle \right)$,  
$1/\sqrt{2}\left( \vert 010 \rangle \pm \vert 101 \rangle \right)$ and  
$1/\sqrt{2}\left( \vert 011 \rangle \pm \vert 100 \rangle \right)$. 
Note that these states are invariant under the group generated by $g_1 = X^{(1)}X^{(2)}\dots X^{(n)}$ and $g_i = X^{(1)}X^{(i)}$ for $2\le i \le n$. This group is local unitary equivalent to the group corresponding to the star graph as shown in Fig.~\ref{fig:graph}, were each node is connected to node one.

A three-qubit state diagonal in the GHZ basis is of the form
\begin{equation}
\label{eqn:ghzdiag}
\varrho =
		 \begin{pmatrix}
		 \lambda_0 & & & & & & & \mu_0 \\
		  & \lambda_1 & & & & & \mu_1 & \\
		 & & \lambda_2 & & & \mu_2 & & \\
		 & & & \lambda_3 & \mu_3 & & &\\
		 & & & \mu_3 & \lambda_3 & & &\\
		 & & \mu_2 & & & \lambda_2 & & \\
		 & \mu_1 & & & & & \lambda_1 & \\
		 \mu_0 & & & & & & & \lambda_0,
		 \end{pmatrix},
\end{equation}
with $\lambda_i, \mu_i \in \mathbb R$.\footnote{GHZ-diagonal states have real $\mu_i$, but we stress that all of our results hold true for states with complex $\mu_i$ as well.} 
A general $n$-qubit GHZ-diagonal state  would have the same shape, with $2^{n-1}$ independent real $\lambda_i$ on the diagonal and corresponding real 
$\mu_i$ on the anti-diagonal. The eigenvalues of these states are $\lambda_i 
\pm \mu_i$, $0\le i < 2^{n-1}$. Hence to be a valid density matrix one needs 
$\lambda_i \ge 0$ and $\vert \mu_i\vert \le \lambda_i$ for all $0 \le i < 
2^{n-1}$.

We now make use of the special structure of this class of states to construct explicit upper and lower bounds, which are valid for both versions of the GMN.

\noindent
{\bf Lemma 5.}
{\it
For all GHZ-diagonal $n$-qubit states 
\begin{equation}
N_g(\varrho) \le \max_i \left\{ 0, \vert\mu_i\vert - w_i \right\} = \max_i \left\{ 0, F_i-\frac 1 2 \right\},
\label{eqn:ghzdiagupperbound}
\end{equation}
where $w_i = \sum_{k\not = i} \lambda_k$ and $F_i = \braket{\psi_i}{\varrho\vert \psi_i}$ denotes the fidelity with the GHZ-basis state $\ket{\psi_i}$. $N_g(\varrho) \le \max_i {0,\vert \mu_i \vert -w_i}$ also holds for the slightly more general case with complex $\mu_i$ on the anti-diagonal.}

{\it Proof.}
We will prove the statement for the three-qubit case, 
a generalization is straightforward. First, consider 
the case where the right-hand-side of Eq.~(\ref{eqn:ghzdiagupperbound}) is nonzero. Without loss of generality one can assume that the maximum is achieved for $i=0$ and thus we have $\vert \mu_0 \vert \ge \sum_{k = 1}^3 \lambda_k$.
Let $p_k = \lambda_k/(\sum_{k = 1}^3 \lambda_k)$ for $1\le k \le 3$, then $\sum_i p_i = 1$ and
\begin{equation}
\label{eqn:pkmugelambda} p_k \vert \mu_0 \vert \ge \lambda_k.
\end{equation}
>From the positivity of $\varrho$ it follows that $\vert \mu_i \vert\le \lambda_i$ and so
\begin{equation}
\label{eqn:pklambdagemu} p_k \lambda_0 \ge p_k \vert \mu_0 \vert \ge \lambda_k \ge \vert \mu_k \vert.
\end{equation}
Using these weights one decomposes $\varrho$ into a convex combination $\varrho = \sum_{k\not = 0}\tilde{p}_k \varrho_k$ with
\begin{eqnarray}
		\varrho_1 = \frac{1}{\tilde{p}_1}
		 \begin{pmatrix}
		 p_1\lambda_0 & & & & & & & p_1\mu_0 \\
		 & \lambda_1 & & & & & \mu_1 & \\
		 & & 0 & & & 0 & & \\
		 & & & 0 & 0 & & &\\
		 & & & 0 & 0 & & &\\
		 & & 0 & & & 0 & & \\
		 & \mu_1^* & & & & & \lambda_1 & \\
		 p_1\mu_0^* & & & & & & & p_1\lambda_0,
		 \end{pmatrix},
\end{eqnarray}
\begin{eqnarray}		 
		\varrho_2 =  \frac{1}{\tilde{p}_k}
		 \begin{pmatrix}
		 p_2\lambda_0 & & & & & & & p_2\mu_0 \\
		 & 0 & & & & & 0 & \\
		 & & \lambda_2 & & & \mu_2 & & \\
		 & & & 0 & 0 & & &\\
		 & & & 0 & 0 & & &\\
		 & & \mu_2^* & & & \lambda_2 & & \\
		 & 0 & & & & & 0 & \\
		 p_2\mu_0^* & & & & & & & p_2\lambda_0,
		 \end{pmatrix},
		 \end{eqnarray}
\begin{eqnarray}
		\varrho_3 =  \frac{1}{\tilde{p}_k}
		 \begin{pmatrix}
		 p_3\lambda_0 & & & & & & & p_3\mu_0 \\
		 & 0 & & & & & 0 & \\
		 & & 0 & & & 0 & & \\
		 & & & \lambda_3 & \mu_3 & & &\\
		 & & & \mu_3^* & \lambda_3 & & &\\
		 & & 0 & & & 0 & & \\
		 & 0 & & & & & 0 & \\
		 p_3\mu_0^* & & & & & & & p_3\lambda_0,
		 \end{pmatrix}
\end{eqnarray}
and $\tilde{p}_k = 2 (p_k \lambda_0+\lambda_k)$.

To calculate the upper bound resulting from this decomposition
we first have to compute the action of partial transposition 
with respect to subsystem $A$, $B$ and $C$ on three-qubit GHZ-diagonal
state. One directly sees that the partial transposition permutes 
the anti-diagonal elements. In the three-qubit case transposition 
of the first qubit exchanges $\mu_0 \leftrightarrow \mu_1$, $\mu_2 \leftrightarrow \mu_3$ and the corresponding conjugate pairs. 
The partial transposition of the second qubit exchanges 
$\mu_0 \leftrightarrow \mu_2$, $\mu_1 \leftrightarrow \mu_3$ 
and conjugate pairs and the partial transposition on the last 
qubit exchanges $\mu_0 \leftrightarrow \mu_3^*$, $\mu_1 \leftrightarrow \mu_2^*$ and conjugate pairs. So $\tilde{p}_k\varrho_k^{T_k}$ has the following four non-zero eigenvalues
\begin{eqnarray}
\left\{ p_k\lambda_0+\vert \mu_k\vert, p_k\lambda_0-\vert \mu_k \vert, \lambda_k+p_k\vert \mu_0 \vert, \lambda_k-p_k\vert\mu_0\vert \right\}.
\end{eqnarray}
Taking into account Eqs.~(\ref{eqn:pkmugelambda}) and 
(\ref{eqn:pklambdagemu}) the only non positive eigenvalue is $ \lambda_k-p_k\vert\mu_0\vert$ and thus $\tilde{p}_k N_k(\varrho_k)=p_k\vert \mu_0 \vert -\lambda_k$. This results in the conjectured upper bound
\begin{equation}
 N_g(\varrho) \le \sum_{k=1}^3 p_k N_k(\varrho_k) =  \sum_{k=1}^3 p_k\vert \mu_0 \vert -\lambda_k = \vert \mu_0 \vert - \sum_{k=1}^3\lambda_k.
\label{eqn:proofupperbound}
\end{equation}
This bound can be rewritten as $ \vert \mu_0 \vert - \sum_{k=1}^3\lambda_k = \vert \mu_0 \vert +\lambda_0 - \frac 1 2$, since $\trace{\vr} = 2\sum_{k=0}^3 \lambda_k = 1$. If we then use the fidelities $F_0 =  \braket{\psi_0}{\varrho\vert \psi_0} = \lambda_0+\mu_0$ and $F_1 =  \braket{\psi_1}{\varrho\vert \psi_1} = \lambda_0-\mu_0$ with $\ket{\psi_0} = 1/\sqrt{2}(\vert 000\rangle + \vert 111 \rangle)$ and $\ket{\psi_1} = 1/\sqrt{2}(\vert 000\rangle - \vert 111 \rangle)$, then $\vert \mu_0 \vert +\lambda_0 = \max\left\{ F_0,F_1 \right\}$ for real $\mu_0$, which proves the alternative expression $N_g(\varrho) \le \max_i \left\{ 0, F_i-\frac 1 2 \right\}$ in Eq.~(\ref{eqn:ghzdiagupperbound}).

If, on the other hand, the right-hand-side of Eq. (\ref{eqn:ghzdiagupperbound}) is zero, then $\vert \mu_0 \vert \le \sum_{k \not = 0} \lambda_k$, which is known to be a necessary and sufficient criterion for biseparability \cite{GuhneSeevinck} and for all biseparable states $\varrho$, $N_g(\varrho)=0$.
\qed

{\bf Lemma 6.}
{\it
Consider a $n$-qubit GHZ-diagonal state $\varrho$ then there 
exists a fully decomposable witness $\mathcal W$ satisfying 
the properties in Eq.~(\ref{eqn:sdpgenneg}), such that 
\begin{equation}
 N_g(\varrho) \ge -\trace ( \mathcal W \varrho) = \min_i \left\{ 0, \vert\mu_i\vert - w_i \right\} = \max_i \left\{0, F_i -\frac 1 2\right\},
\label{eqn:ghzdiaglowerbound}
\end{equation}
where $w_i = \sum_{k\not = i} \lambda_k$. $N_g(\varrho) \ge \max_i {0,\vert \mu_i \vert -w_i}$ also holds for the slightly more general case with complex 
$\mu_i$ on the antidiagonal.}

{\it Proof.}
As in the last proof we consider the three-qubit case. Without loss of generality the minimum in inequality (\ref{eqn:ghzdiaglowerbound}) 
is achieved for $i=0$.
Then the position of $\mu_0$ in $\varrho$ in the computational basis is given by the tuple $(000,111)$. From this tuple we construct the witness $\mathcal W = \frac 1 2 \eins - \ketbra{\phi}$, with $\ket{\phi} = \frac{1}{\sqrt{2}}\left( \ket{000}+\ket{111} \right)$. In the more general case, where the $\mu_i\in \mathbb C$ one would insert an additional phase $e^{i \arg{\mu_0}}$ in front of $\ket{111}$.

The witness is of the form of $\varrho$ as in Eq.~(\ref{eqn:ghzdiag}). From
the discussion in the proof of Lemma 5 it follows that
$\mathcal W^{T_m}\ge 0$. Hence, $\mathcal W$ is fully decomposable with $P_m=0$ and $Q_m = W^{T_m}$. In case the minimum equals zero the witness is given by $\mathcal W=0$, $P_m=0$ and $Q_m=0$. This proves the claim.
\qed

In summary, we have:

\noindent
{\bf Corollary 7.}
{\it
For all GHZ-diagonal $n$-qubit states $\varrho$}
\begin{equation}
N_g(\varrho) = \max_i \left\{ 0, \vert\mu_i\vert - w_i \right\} = \max_i \left\{0, F_i -\frac 1 2\right\},
\label{eqn:ghzdiagclosedformula}
\end{equation}
{\it where $F_i = \braket{\psi_i}{\varrho\vert \psi_i}$ denotes the fidelity with the GHZ-basis state $\psi_i$. $N_g(\varrho) = \max_i {0,\vert \mu_i \vert -w_i}$ holds also true for the slightly more general case with complex $\mu_i$ on the anti-diagonal.}

Three remarks are in order at this point. First, for general states the right hand side of Eq.~(\ref{eqn:ghzdiagclosedformula}) still gives a lower bound on the renormalized GMN since every state can be transformed into a GHZ-diagonal by means of local operations only. Thus the analytic expression in Eq.~(\ref{eqn:ghzdiagclosedformula}) might be used to estimate genuine multiparticle entanglement also for a general state. Second, 
we calculated
the value for the renormalized GMN $N_g(\varrho)$ as defined in
Eq.~(\ref{eqn:sdpgenneg}), but the same result holds for the 
GMN $\tilde N_g(\varrho)$ according to Eq.~(\ref{eqn:sdpgennegold}). 
This is because
the witness constructed in the proof of Lemma 6 fulfils also
the conditions of Eq.~(\ref{eqn:sdpgennegold}) and Lemma 5 
delivers an upper bound due to Corollary 4. Third, note that 
the analytic formula we found coincides with the maximal 
violation of the biseparability criteria derived in 
Ref.~\cite{GuhneSeevinck}.

Further, it is interesting to compare our expression for the 
GMN to the analytic formula of 
the genuine multiparticle concurrence (GME-concurrence)
\cite{GME-concurrence} for GHZ-diagonal $n$-qubit states \cite{GMECX}. The GME-concurrence is defined as follows: 
For a  bipartite pure state $\ket{\psi}$ the concurrence
is given by $C^{A|B}(\psi)= \sqrt{2[1-Tr(\vr_A^2)]}$ where $\vr_A$
is the reduced state on Alice's system. For a pure multipartite
state $\ket{\phi}$, the genuine multiparticle concurrence is defined 
as the minimum of the bipartite concurrences $C^{\rm gme}(\phi)=
\min_{m \vert \bar m} C^{m \vert \bar m}(\phi)$, 
minimized over all bipartitions. 
Finally, for mixed states the measure is given by the convex roof 
construction
\be
C^{\rm gme}(\vr) = \min_{\vr=\sum_k p_k \ketbra{\phi_k}}
\sum_k p_k C^{\rm gme}(\phi_k).
\label{gme-conv}
\ee
Note that contrary to the mixed convex roof optimization in 
Eq.~(\ref{eqn:genuinenegativity}) here only pure state 
decompositions are involved. This pure state convex roof, however, 
is in general nearly impossible to compute and therefore one relies 
on lower bounds for practical applications.

The GME-concurrence has been computed for GHZ-diagonal states \cite{GMECX} and one observes that up to a factor 
of two both expressions coincide. There are, however, deeper 
connections: for a pure multiqubit state one has 
\be
N_g(\phi) \leq C^{\rm gme}(\phi).
\ee
This follows directly from known relations between the bipartite
negativity and the bipartite concurrence \cite{chenalbeverio}. 
Since the GMN can be
defined via the {\it mixed} convex roof, which is an optimization over
a larger set than the pure convex roof of the GME-concurrence, the 
general bound
\be
N_g(\vr) \leq C^{\rm gme}(\vr).
\ee
holds for all mixed states. Therefore, Theorem 2 provides a
way to obtain lower bounds on the GME-concurrence via semidefinite
programming or analytical calculations. We will see in the next section, 
that the value of the $N_g(\vr)$ is much more sensitive 
to entanglement than the known lower bounds on the GME-concurrence.

\subsection{Four-qubit cluster-diagonal states}

In this section we consider the linear graph having four 
vertices as shown in Fig.~\ref{fig:graph}. The corresponding 
stabilizing operators as defined in Eq.~(\ref{eqn:graphstabilizer}) 
are given by
\begin{eqnarray}
\nonumber &g_1 = XZ\eins\eins, \ \ \ \ \ \ g_2 = ZXZ\eins,\\
&g_3 = \eins ZXZ \mbox{ and } g_4 = \eins\eins ZX.
\label{eqn:clusterstabilizer}
\end{eqnarray}
As already discussed, the corresponding graph state basis is 
given by $\ket{++++}$, $\ket{+++-}$, $\dots$, $\ket{----}$. We will 
also write this states as $\ket{ijlk}$, with $i,j,l,k \in \{+,-\}$ 
and denote by $\bar{k}$ the complement of $k$. Then we consider the
graph-diagonal state
\begin{equation}
\varrho = \sum_{i,j,k,l} F_{ijkl} \ketbra{ijkl}
\label{eqn:clusterdiag}
\end{equation}
and wish to compute the GMN for these states. Note that in the literature
the four-qubit cluster state is often defined via the local unitary 
equivalent stabilizing operators $\tilde g_1 = ZZ\eins\eins$, 
$\tilde g_2 = XXZ\eins$, $\tilde g_3 = \eins Z XX$ and 
$\tilde g_4 = \eins\eins ZZ$. Then the corresponding eigenstate 
to eigenvalue $+1$ on all $\tilde g_i$ is the familiar
cluster state 
\be
\ket{CL}=\frac{1}{2}(\ket{0000}+\ket{1100}+\ket{0011}-\ket{1111})
\ee
in the computational basis.

For discussing genuine multiparticle entanglement of cluster
diagonal states, the following two classes of witnesses 
\begin{eqnarray}
\label{eqn:clusterwone} &\mathcal W_{\alpha\beta\gamma\delta} = \frac{\eins}{2} - \ketbra{\alpha\beta\gamma\delta} -\frac 1 2 \sum_{i,j} \ketbra{\bar{\alpha} i j \bar{\delta}},\\
\label{eqn:clusterwtwo} &\mathcal W_{\alpha\beta\gamma\delta\mu\nu} =  \frac{\eins}{2} - \ketbra{\alpha\beta\gamma\delta} - \ketbra{\bar{\alpha} \mu\nu \bar{\delta}},
\end{eqnarray}
have turned our to be useful.  It has been shown that these witnesses
provide necessary and sufficient criteria to detect genuinely 
multiparticle entanglement in these states \cite{PhysRevA.84.052319}. 
Based on these, Chen et al. \cite{PhysRevA.87.012322} provided 
a closed formula for the genuine multiparticle relative entropy 
of entanglement as entanglement monotone. Here we provide a closed 
formula for the GMN for four-qubit cluster-diagonal states. 

In terms of the fidelity the expectation values of the witnesses in 
Eqs.~(\ref{eqn:clusterwone}) and (\ref{eqn:clusterwtwo}) read
\bea
\nonumber 
\trace({\mathcal W_{\alpha\beta\gamma\delta}\varrho}) &=& -F_{\alpha\beta\gamma\delta} + \frac 1 2 \sum_{ij} F_{\alpha ij \bar{\delta}} + F_{\bar{\alpha} ij \delta} + F_{\alpha ij \delta},
\\
\label{eqn:clusterwexp} 
\trace({\mathcal W_{\alpha\beta\gamma\delta\mu\nu} \varrho}) &=& -F_{\alpha\beta\gamma\delta} - F_{\bar{\alpha}\mu\nu\bar{\delta}} + \frac 1 2 .
\eea
First we note that all of these witnesses can be used to 
bound the GMN from below:

\noindent
{\bf Lemma 8.}
{\it
For all partitions $m\vert  \bar m$ of the four partied system $ABCD$ there exists $0\le Q_m, \tilde Q_m \le \eins$, such that}
\begin{eqnarray}
\mathcal W_{\alpha\beta\gamma\delta} = Q_m^{T_m}  & \mbox{ and } &
\mathcal W_{\alpha\beta\gamma\delta\mu\nu} = \tilde{Q}_m^{T_m}.
\end{eqnarray}

{\it Proof.}
One can easily compute the eigenvalues of $Q_m = \mathcal W_{++++}^{T_m}$ and $\tilde{Q}_m = \mathcal W_{++++ij}^{T_m}$ for all $i,j\in\left\{ +,- \right\}$ to be $0$ and $\frac 1 2$. For any other witness $\mathcal W_{\alpha\beta\gamma\delta}$ ($\mathcal W_{\alpha\beta\gamma\delta\mu\nu}$) there exists a local unitary transformations, such that $\mathcal W_{++++}$ ($\mathcal W_{++++ij}$) transforms into it. 
\qed

Taking into account that each of the above witnesses gives a lower bound on the GMN [see Eq. (\ref{eqn:lowerbound})] we have that
\begin{equation}
 -\min_{\alpha,\beta,\gamma,\delta,\mu,\nu} \left\{ \trace(\mathcal W_{\alpha\beta\gamma\delta}\varrho)  \right\} \cup \left\{ \trace(\mathcal W_{\alpha\beta\gamma\delta\mu\nu}\varrho) \right\} \cup \left\{ 0 \right\} \le N_g(\rho),
\label{eqn:clusterlowerbound}
\end{equation}
for all  four-qubit cluster-diagonal states.
As done for the GHZ-diagonal states we can use specific decompositions [see Eq. (\ref{eqn:genuinenegativity})] to construct upper bounds on the GMN, which result in an analytic formula for all four-qubit cluster states.

\noindent
{\bf Theorem 9.}
{\it
Let $\varrho = \sum_{\alpha\beta\gamma\delta} F_{\alpha\beta\gamma\delta} \ketbra{\alpha\beta\gamma\delta}$ be diagonal in the cluster graph basis, then the GMN of that state is given by
\begin{equation}
N_g(\varrho) = - \min_{\alpha,\beta,\gamma,\delta,\mu,\nu}  \left\{ \trace(\mathcal W_{\alpha\beta\gamma\delta}\varrho)  \right\} \cup \left\{ \trace(\mathcal W_{\alpha\beta\gamma\delta\mu\nu}\varrho) \right\} \cup \left\{ 0 \right\}.
\label{eqn:clustergenneg}
\end{equation}
This means that effectively the largest violation of the witnesses
in Eqs.~(\ref{eqn:clusterwone}) and (\ref{eqn:clusterwtwo}) gives the
value of the GMN.}

The proof of this Theorem is given in \ref{sec:AB}.

\begin{figure}[t]
\begin{center}
\includegraphics[width=0.7\textwidth]{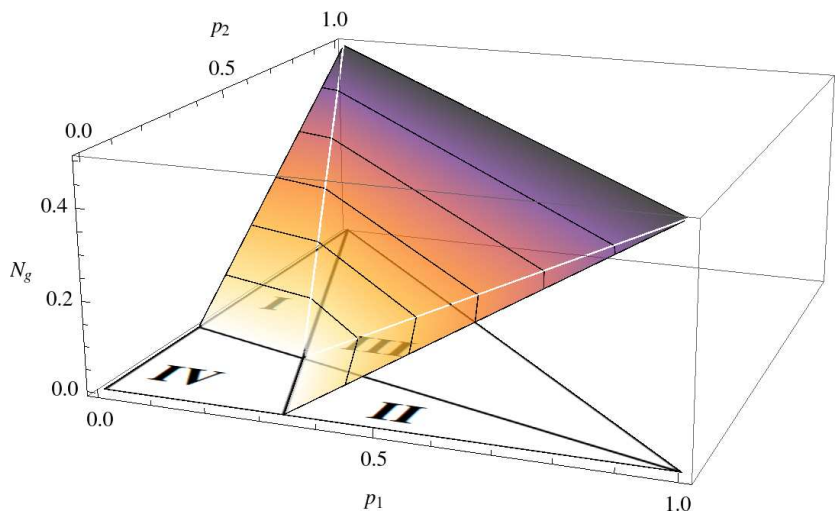}
\end{center}
\caption{The GMN for a two-parameter family of cluster-diagonal states, 
given by $\varrho = p_1 \ketbra{+\hspace{-0.12em}+\hspace{-0.12em}+\hspace{-0.12em}+} +p_2 \ketbra{-\hspace{-0.12em}+\hspace{-0.12em}+\hspace{-0.12em}-} + (1-p_1-p_2) \frac 1 2 (\sigma_1+\sigma_2)$, with biseparable $\sigma_1 = \frac 1 2(\ketbra{+\hspace{-0.12em}+\hspace{-0.12em}-\hspace{-0.12em}+}+\ketbra{+\hspace{-0.12em}-\hspace{-0.12em}+\hspace{-0.12em}+})$ and $\sigma_2 = \frac 1 2 (\ketbra{-\hspace{-0.12em}+\hspace{-0.12em}-\hspace{-0.12em}-}+\ketbra{-\hspace{-0.12em}-\hspace{-0.12em}+\hspace{-0.12em}-}) $.  In the regions I to III the GMN $N_g$ 
is given by the negative expectation value of
different witnesses (see Eq.~(\ref{eqn:clustergenneg})). 
In I it corresponds to $-\trace(\mathcal W_{-++-}\varrho) = 
\frac 1 4 (p_1 +3 p_2 - 1)$, 
in II it is given by $-\trace(\mathcal W_{++++++}
\varrho) = p_1+p_2 - \frac 1 2$ and in region III it 
is $-\trace(\mathcal W_{++++}\varrho) = \frac 1 4 (3p_1+p_2-1)$. 
In the remaining region IV the state is biseparable.}
\label{fig:cluster}
\end{figure}

Let us discuss some examples.
For the graph state mixed with white noise $\varrho = p\ketbra{++++} + (1-p)\eins/16$ we obtain 
$N_g(\varrho) = \max(\frac{13p -5}{16},0)$, which gives the exact threshold $p>5/13$ for genuine multiparticle entanglement~\cite{PhysRevA.84.052319}. 
In Fig.~\ref{fig:cluster} a two-parameter family  is shown, which
is is genuine multiparticle entangled in three regions and 
biseparable in one. In each of the regions a different optimal 
witness gives the GMN.

Finally, we compare our results to computable lower bounds on the 
GME-concurrence introduced in Refs.~\cite{GME-concurrence, PhysRevA.83.040301} and general lower bounds on the linear entropy 
based genuine multiparticle entanglement measure in Ref.~\cite{PhysRevA.86.022319}. To compare the performance we calculate the value $p$ down to which  $\varrho = p\ketbra{++++} + (1-p)\eins/16$ is still detected as genuine multiparticle entangled. 
Using the general bound of the GME-concurrence in 
Ref.~\cite{GME-concurrence} we found that even the pure four-qubit cluster-diagonal state is not detected. Using instead a set of inequalities 
build to detect genuine multiparticle entanglement in $n$-qubit Dicke states \cite{PhysRevA.83.040301} we found $\varrho$ to be 
detected for $p>0.982$.\footnote{Note that we applied local 
filters to the state $\varrho \mapsto N (F_A^\dagger\otimes F_B^\dagger \otimes F_C^\dagger \otimes F_D^\dagger \varrho F_A\otimes F_B \otimes F_C \otimes F_D)$ to enhance its detectability,
where $N$ is a normalization and the $F_i$ are linear maps on the single qubit systems.}
A better detection was achieved with the general lower bound on the genuine multiparticle entanglement measure given by Theorem 1  from 
Ref.~\cite{PhysRevA.86.022319}. We found that 
the state $\varrho$ was detected as genuine multiparticle entangled for $p>7/15\approx0.47$, which is 
closer to the exact threshold $p>5/13\approx 0.38$ but not the
exact value.
We can therefore conclude that although the analytic formula 
for the GME-concurrence is equivalent to ours for $n$-qubit 
GHZ-diagonal states, the lower bounds for four-qubit 
cluster-diagonal states do not match our analytic results.

\section{Conclusions}
In conclusion we have shown that the renormalized genuine multiparticle 
negativity can be expressed in two equivalent ways: As an optimization over 
suitable normalized fully decomposable witnesses as given by 
Eq.~(\ref{eqn:sdpgenneg}) and as mixed convex roof of the minimal 
bipartite negativity as given by Eq.~(\ref{eqn:genuinenegativity}). As a 
direct consequence of these equivalent definitions there are naturally 
arising lower and upper bounds, which we used to obtain an exact algebraic 
prescription of the genuine multiparticle negativity for the $n$-qubit 
GHZ-diagonal and four-qubit cluster-diagonal states. These analytic 
expressions can also be used to obtain lower bounds on the genuine multiparticle negativity for arbitrary $n$-qubit states.

There are several questions arising, which one might investigate in the future. 
First, since the scheme to obtain the analytic expression is quite general it 
should be possible to find closed expressions for other highly symmetric state 
families like other graph-diagonal states \cite{Graphstatereview} or  
states with $U\otimes U \otimes U$ symmetry \cite{wernerstates}.

Second, it would be desirable to obtain an operational interpretation for the 
genuine multiparticle negativity. As the bipartite logarithmic negativity is 
the upper bound for distillable entanglement \cite{VidalWerner} one may speculate 
that our monotone is connected to the distillation rate of genuine multiparticle 
entangled states. Also, the multiparticle negativity may be related to different
entanglement classes in the multiparticle case and the dimensionality of multiparticle
entanglement \cite{marcusdim}.

Finally, recall that the shareability of quantum correlations among many 
parties is limited and these restrictions 
are known as monogamy relations \cite{coffman, koashi, negmon}. 
For example, for a three-qubit system the bipartite entanglement of 
the splitting $A\vert BC$ as measured by the concurrence is given 
by the entanglement in the reduced marginals plus the three tangle 
$\tau_3$ as a genuine tripartite contribution, 
$ C_{A\vert BC}^2 = C_{AB}^2+C_{AC}^2+ \tau_3$ \cite{coffman}. It would be
very interesting to derive similar relations for the genuine multiparticle
negativity.

\ack
We thank Marcus Huber for helpful discussions. This work  
has been supported by the EU (Marie Curie CIG 293993/ENFOQI),
the BMBF (Chist-Era Project QUASAR), the FQXi Fund
(Silicon Valley Community Foundation) and the DFG.

\appendix
\setcounter{section}{0}

\section{Proof of Theorem 2}
\label{sec:AA}
To start, we recall some notions of semidefinite programing \cite{SDP}.
The primal problem of an SDP reads
\begin{eqnarray}
\inf_{\vec x}&\ \vec{c}^T \vec x\\
\nonumber	\mbox{s.t. }& F(\vec x) = F_0 + \sum_i x_iF_i \ge 0,
\end{eqnarray}
where $\vec c,\vec x \in \mathbb R ^n$ and $F_i=F_i^\dagger \in \mathbb C^{m \times m}$.
The scalar product ${\vec c}^T \vec x$ is the linear function to minimize and 
the linear matrix inequality $F(\vec x)\ge 0$, understood in terms of positive 
semi-definiteness holds all the constraints of the optimization. The dual 
problem to this primal problem is given by
\begin{eqnarray}
\label{eqn:dualprob}	\sup_{Z}& \left[ - \trace\left( F_0 Z \right) \right]\\
\nonumber	\mbox{s.t. }&  \trace\left( F_i Z  \right) = c_i \ \mbox{ for all } i=1,\dots,n,\\
\nonumber       & Z\ge 0.
\end{eqnarray}
A point $\vec x$ or $Z$ is called feasible if it meets the constraints of 
the primal $F(\vec x) \ge 0$ or dual respectively $Z\ge 0$ and $\trace(F_iZ)=c_i$. 
For any pair of feasible points both problems are connected to each 
other via $- \trace{F_0Z} \le \vec{c}^T \vec x$. Moreover, if at least one of the problems is strictly feasible, i.e. that either a primal point $\vec x$ satisfying $F(\vec x)>0$ or a dual point $Z$ satisfying $Z>0$ and $\trace(F_iZ)=c_i$ exists, Theorem 3.1 in Ref.~\cite{SDP} 
ensures that both problems yield the same optimum 
$\sup_Z \left\{ -\trace(F_0 Z) \vert Z\ge 0\text{ and }\trace(F_iZ)=c_i \right\} = \inf_{\vec x} \left\{ \vec c^T \vec x \vert F(\vec x) \ge 0 \right\}$.

The idea of the proof goes as follows. Using our renormalized GMN 
as given by Eq. (\ref{eqn:sdpgenneg}) as the primal problem of a SDP we show 
that the corresponding dual problem is given by the left-hand-side of 
Eq.~(\ref{eqn:genuinenegativity}). Equality then follows by showing strict 
feasibility of the primal problem.
We start the proof by rewriting the semidefinite program 
(\ref{eqn:sdpgenneg}) as
\begin{eqnarray}
\label{eqn:sdpgennegproof}	
N_g(\varrho) &=&- \inf\ \trace\left( \varrho \mathcal W \right)\\
\nonumber	
\mbox{ s.t. } && 0\le P_m, \\
\nonumber	
&&0\le (\mathcal W-P_m)^{T_m} \le\eins\  \mbox{ for all partitions } m.
\end{eqnarray}
For the sake of readability we write down the proof by assuming a quantum 
system composed of three parts $A$, $B$ and $C$. A generalization to 
larger particle numbers is straightforward.

We choose a Hermitian operator basis  $\sigma_i $, $i=1,\dots,K$, 
such that $\trace\left( \sigma_i\sigma_j \right) = \delta_{ij}$. 
In this basis $\varrho = \sum_i \varrho^{(i)}\sigma_i$, 
$\mathcal W= \sum_i w^{(i)}\sigma_i$ and 
$P_m = \sum_i p_m^{(i)}\sigma_i$ for $m \in \left\{ A,B,C \right\}$. 
We gather the components of this decompositions in the vectors
\begin{eqnarray}
\vec x_w = \left( w^{(1)}, \dots. , w^{(K)} \right),
&\mbox{ } &
\vec x_m = \left( p^{(1)}_m, \dots. , p_m^{(K)} \right),
\nonumber
\\
\vec c_w = \left( \varrho^{(1)}, \dots. , \varrho^{(K)} \right),
&\mbox{ }  &
\vec c_m = \vec 0,
\end{eqnarray}
where $\vec x_w$ are the coefficients of $\mathcal W$, $\vec x_m$ are the coefficients of $P_m$ and $\vec c_w$ as well as the $\vec c_m$ characterize 
parts of the optimization goal. If we merge these vectors into
\begin{equation}
	\label{eqn:xc} \vec x =\left(\vec x_w,\vec x_A,\vec x_B,\vec x_C \right)
	\mbox{ and }\vec c=\left(\vec c_w,\vec c_A,\vec c_B,\vec c_C \right),
\end{equation}
we can rewrite the SDP (\ref{eqn:sdpgennegproof}) as
\begin{eqnarray}
	\label{eqn:standardsdp}	&-\inf_{\vec x}\ \vec{c}^T \vec x\\
\nonumber	\mbox{s.t. }& F(x) = F_0 + \sum_i x_iF_i \ge 0,
\end{eqnarray}
where $F(\vec x)$ has the following block diagonal form
\bea
F(\vec x) &= & 
diag ( P_A,P_B,P_C \;\vert\; (\mathcal W-P_A)^{T_A},(\mathcal W-P_B)^{T_B},(\mathcal W-P_C)^{T_C} \;\vert\;
\nonumber
\\
&&\eins-(\mathcal W-P_A)^{T_A},\eins-(\mathcal W-P_B)^{T_B},\eins-(\mathcal W-P_C)^{T_C}).
\eea
Here the vertical lines ``$\vert$'' are introduced for notational convenience
in order to to split the block diagonal matrix $F(\vec x)$ into three parts, 
each of which consists of three sub blocks. The first represents 
the constraint $0\le P_m$, the second ensures $0\le(\mathcal W-P_m)^{T_m}$ 
(equivalent to $0\le Q_m$) and the last one bounds 
$(\mathcal W-P_m)^{T_m}\le \eins$ (equivalent to $Q_M\le \eins$) for 
all $m \in \left\{ A,B,C \right\}$. According 
to 
$F(\vec x) = F_0 + \sum_i x_iF_i = F_0 + \sum_j (\vec x_w)_jF_{w,j} + \sum_m \sum_j (\vec x_m)_jF_{m,j}$, we have:
\begin{eqnarray}
F_0 &= diag\left(0,0,0 \;\vert\; 0,0,0 \;\vert\; \eins,\eins,\eins \right),
\nonumber
\\
F_{w,j} &= diag \left( 0,0,0 \;\vert\; \sigma_j^{T_A},\sigma_j^{T_B},\sigma_j^{T_C}\;\vert\; -\sigma_j^{T_B},-\sigma_j^{T_B},-\sigma_j^{T_C}\right),
\nonumber
\\
F_{A,j} &= diag\left( \sigma_j,0,0 \;\vert\; -\sigma_j^{T_A},0,0 \;\vert\; \sigma_j^{T_A},0,0 \right),
\nonumber
\\
F_{B,j} &= diag\left( 0,\sigma_j,0 \;\vert\; 0,-\sigma_j^{T_B},0 \;\vert\; 0,\sigma_j^{T_B},0 \right),
\nonumber
\\
F_{C,j} &= diag\left( 0,0,\sigma_j \;\vert\; 0,0,-\sigma_j^{T_C} \;\vert\; 0,0,\sigma_j^{T_C} \right).
\end{eqnarray}

The dual problem as given in Eq.~(\ref{eqn:dualprob}) involves the 
calculation of $\trace\left( F_0 Z \right)$ and $\trace\left( F_i Z \right)$. 
To do so we can make use of the block-diagonal structure of the $F_i$ and 
write the corresponding diagonal blocks of $Z$ into a new block-diagonal matrix
\begin{equation}
Z_{bd} =diag\left( Z_A,Z_B,Z_C \;\vert\; Z_A^+,Z_B^+,Z_C^+ \;\vert\; Z_A^-,Z_B^-,Z_C^- \right).
\label{eqn:bdpartofZ}
\end{equation}
Note that the positivity of $Z$ ensures the positivity of each block in $Z_{bd}$. On the other hand if the blocks in $Z_{bd}$ are positive so is $Z_{bd}$.
We can now evaluate $-\trace\left( F_0Z \right) = -\trace\left( F_0Z_{bd} \right)$ 
to write down the dual objective
\begin{equation}
-\sup_{Z \ge 0}\left[  -\trace\left( Z_A^- \right)-\trace\left( Z_B^- \right)-\trace\left( Z_C^- \right) \right] = \inf_{Z\ge 0} \sum_m \trace\left( Z_m^- \right).
\label{eqn:dualobjective}
\end{equation}
The constraints $\trace\left( F_iZ \right) = c_i$ can be evaluated similarly and split up into two types $\trace\left( F_{w,j}Z \right) = \varrho_j$ and $\trace\left( F_{m,j}Z \right) = 0$. In detail, they read
\begin{eqnarray}
\label{eqn:constrainttwo} 
\sum_m \trace\left( \sigma_j^{T_m}Z_m^+ \right) - \trace\left( \sigma_j^{T_m}Z_m^- \right)& =& \varrho^{(j)},\\
\label{eqn:constraintone} \trace\left( \sigma^j Z_m \right)-\trace\left( \sigma_j^{T_m}Z_m^+ \right) + \trace\left( \sigma_j^{T_m}Z_m^- \right) &=& 0,
\end{eqnarray}
with $m\in \left\{ A,B,C \right\}$. If we multiply Eq.~(\ref{eqn:constraintone}) 
by $\sigma_j$ and sum over all $j$ it immediately follows that
\begin{equation}
\label{eqn:decompZ0m} Z_m = {Z_m^+}^{T_m} - {Z_m^-}^{T_m}.
\end{equation}
Eq.~(\ref{eqn:constrainttwo}) multiplied by $\sigma_j$ and summed over $j$ together with Eq.~(\ref{eqn:decompZ0m}) yields
\begin{equation}
\label{eqn:convexcomb} \varrho = \sum_m {Z_m^+}^{T_m}-{Z_m^-}^{T_m} = \sum_m Z_m.
\end{equation}
Actually, the dual problem optimizes in state space, which can be made apparent 
by introducing the following notation. Let $p_m = \trace(Z_m)$ and 
$\varrho_m = Z_m / \trace(Z_m)$, then the constraint given by 
Eq.~(\ref{eqn:convexcomb}) corresponds to an optimization over all 
possible convex combinations $\varrho = \sum_m p_m \varrho_m$ of mixed 
quantum states $\varrho_m$. By introducing $\varrho_m^\pm = Z_m^\pm / \trace(Z_m)$ 
the constraint given by Eq.~(\ref{eqn:decompZ0m}) can be rewritten 
as $\varrho_m^{T_m} = \varrho_m^+-\varrho_m^-$. The dual problem 
is then given by
\begin{eqnarray}
\nonumber \inf & \ p_A \trace\left( \varrho_A^- \right) + p_B\trace\left( \varrho_B^- \right) + p_C\trace\left( \varrho_C^- \right)\\
\nonumber \mbox{s.t.}	& \varrho = p_A\varrho_A+p_B\varrho_B + p_C\varrho_C \mbox{ is a decomposition of }\varrho\mbox{ and }\\
	& \varrho_m^{T_m} = \varrho_m^+ - \varrho_m^- \ \mbox{ for all }m\in \left\{ A,B,C \right\}
\label{eqn:dualgenneg}	 \mbox{with } \varrho^\pm \ge 0,
\end{eqnarray}
which means that given a density matrix $\varrho$ one minimizes 
$ \sum_m p_m \trace\left( \varrho_m^- \right)$ over all 
decompositions $\varrho = \sum_m p_m\varrho_m$ and respective 
splittings of the partial transposition $\varrho_m^{T_m}$ into 
a difference of two positive semidefinite operators 
$\varrho_m^{T_m} = \varrho_m^+ - \varrho_m^-$.

Note that one can split this optimization into two steps. First, one has to 
optimize over all mixed state decompositions of $\varrho$, where each term 
in the decomposition is assigned to a certain bipartition. For a fixed 
decomposition $\varrho = \sum_m p_m\varrho_m$ one still has to minimize 
$\sum_m p_m\ \trace\left( \varrho_m^- \right)$ over 
$\bigcup_m \mathcal N_m$ with 
$\mathcal N_m = \{\varrho_m^\pm\ge 0 \vert \varrho_m^{T_m} = \varrho_m^++\varrho_m^-\}$.
This can be decomposed into the separate minimization of each 
$\trace\left( \varrho_m^- \right)$ over $\mathcal N_m$. If one 
compares these single optimizations to the bipartite 
negativity \cite{VidalWerner} of the respective partition
\begin{equation}
	N_m(\varrho) = \inf\left\{ \trace\left( \varrho^- \right)\vert \varrho^{T_m} = \varrho^+-\varrho^-, \varrho^\pm\ge 0 \right\},
	\label{eqn:negativity}
\end{equation}
then $\min_{\mathcal N_m} \trace\left( \varrho_m^- \right) = N_m(\varrho_m)$. Hence we can rewrite the dual problem as given in Eq.~(\ref{eqn:dualgenneg}) 
by
\begin{eqnarray}
\nonumber \min & \ p_A N_A(\varrho_A) + p_B N_B(\varrho_B) + p_C N_C(\varrho_C),\\
\nonumber \mbox{s.t. }	& \varrho = p_A\varrho_A+p_B\varrho_B + p_C\varrho_C.
\end{eqnarray}
Here we replaced the infimum by a minimum, since one optimizes a continuous function over a closed and bounded set.

To finish this proof we still have to show that the primal 
problem is strictly feasible such that both problems have the same optimal value. We find that 
$\mathcal W = P_m + Q_m^{T_m}$ with $Q_m=P_m = \eins/2 >0$ is a 
strictly feasible point for the primal problem given by Eq.~(\ref{eqn:sdpgenneg}). Hence, the genuine multiparticle 
negativity equals the dual optimization problem
\begin{equation}
N_g(\varrho) = \min_{\varrho = \sum_m p_m\varrho_m} \sum_m p_m N_m(\varrho_m).
\end{equation}
\qed

\section{Proof of Theorem 9}
\label{sec:AB}
Within this proof we will shorten the Dirac notation by 
setting $\ketbra{\alpha\beta\gamma\delta} =\ketbras{\alpha\beta\gamma\delta}$.
First, recall that a four-qubit cluster-diagonal state 
is biseparable, iff the following inequalities are 
satisfied \cite{PhysRevA.84.052319}:
\begin{align}
\label{eqn:clusterbisepone} F_{\alpha\beta\gamma\delta} & \le \frac 1 2 \sum_{ij} F_{\bar{\alpha} ij \delta} + F_{\alpha ij \bar{\delta}} +F_{\alpha ij \delta} \quad \mbox{ and }
\\
\label{eqn:clusterbiseptwo} F_{\alpha\beta\gamma\delta} + F_{\bar{\alpha} \mu\nu \bar{\delta}} & \le \frac 1 2 \sum_{ij} F_{\alpha ij \delta} +F_{\bar{\alpha} ij \delta} + F_{\alpha ij \bar{\delta}} + F_{\bar{\alpha} ij \bar{\delta}}.
\end{align}
Furthermore, Lemma 2 in Ref.~\cite{PhysRevA.84.052319} states 
that the density matrix
\begin{equation}
\vr^{\rm bs} = \frac 1 2 \left( \ketbras{ijkl} + \ketbras{\alpha\beta\gamma\delta} \right)
\label{eqn:clusterbs}
\end{equation}
is biseparable, unless $i\not = \alpha$ and $l \not = \delta$ both hold at the same time.

We now prove that the maximal violation of the relations
(\ref{eqn:clusterbisepone}) and (\ref{eqn:clusterbiseptwo}) 
[this corresponds to the negative of the expectation values 
of some witness in Eqs.~(\ref{eqn:clusterwexp})] is an upper
bound on the GMN. 
There are three cases:

{\it Case one}: None of the inequalities (\ref{eqn:clusterbisepone}) and (\ref{eqn:clusterbiseptwo}) is violated. In this case we already know that the state is biseparable \cite{PhysRevA.84.052319} and hence $N_g(\varrho) = 0$, which coincides with the right-hand-side of (\ref{eqn:clustergenneg}) in this case.

{\it Case two}: The largest violation occurs in inequality (\ref{eqn:clusterbisepone}). We can assume that it occurs 
for $\alpha=\beta=\gamma=\delta=+$, for other indices 
the reasoning is similar. 
Using $\sum_{ijkl} F_{ijkl} = 1$ and the fact 
that all other inequalities are less violated we obtain
\begin{equation}
F_{++++} - \frac 1 2 \sum_{ij} F_{+ ij +} + F_{- ij +} + F_{+ ij -}
\ge F_{++++} + F_{- \mu\nu -} - \frac 1 2,
\label{eqn:ineqoneleineqtwo}
\end{equation}
for $\mu,\nu$ arbitrary. Adding $\frac 1 2 \sum_{ijkl} F_{ijkl} = \frac 1 2$ on both sides yields
\begin{equation}
F_{- \mu\nu -} \le F_{- \mu \bar{\nu} -} + F_{- \bar{\mu} \nu -} + F_{- \bar{\mu} \bar{\nu} -}.
\end{equation}
Now consider the state $\sigma \sim \sum_{\mu\nu} F_{-\mu\nu -} \ketbras{\!\!-\!\! \mu\nu -}$. One can check that $\sigma$ does not violate any of the inequalities (\ref{eqn:clusterbisepone}) and (\ref{eqn:clusterbiseptwo}), so it is biseparable. 
We choose $F_{++++}^{\mathit{ent}} = 2F_{++++} - \sum_{ij} F_{+ ij +} + F_{- ij +} + F_{+ ij -}$, which is two times the violation of 
inequality (\ref{eqn:clusterbisepone}). Then we can decompose $\varrho$ into a genuine multiparticle entangled part with weight $F_{++++}^{\mathit{ent}}$ and a biseparable rest. This rest consists of a convex combination of biseparable states as in Eq. (\ref{eqn:clusterbs}) and the biseparable state $\sigma$
\begin{align}
\nonumber \varrho & =  F_{++++}^{\mathit{ent}} \ketbras{++++}
\;\;
+ \sum_{ij, ij \not = ++} F_{+ ij +} \left( \ketbras{++++} 
+ \ketbras{+ ij +} \right) 
\\
\nonumber 
&+ \sum_{ij} F_{+ ij -} \left( \ketbras{++++} + \ketbras{+ ij -} \right) 
+ \sum_{ij} F_{- ij +} \left( \ketbras{++++} +\ketbras{- ij +} \right) \\
& +\sum_{ij} F_{- ij -} \ketbras{- ij -}.
\end{align}
Since only the first part is not biseparable and the GMN is convex 
$N_g(\sum_m p_m \varrho_m) \le \sum_m p_m N_g(\varrho_m)$, 
we obtain
\be
N_g(\varrho) 
\le F_{++++}^{\mathit{ent}}N_g(\ketbras{++++})\\
= \frac{1}{2}
\big( 2F_{\alpha\beta\gamma\delta} - \sum_{ij} F_{\alpha ij \delta} +F_{\bar{\alpha} ij \delta} + F_{\alpha ij \bar{\delta}} \big) 
\ee
which corresponds to the right-hand 
side of Eq.~(\ref{eqn:clustergenneg})
in this case.

{\it Case three}:
The largest violation occurs in Eq.~(\ref{eqn:clusterbiseptwo}). 
Without loss of generality for $\alpha=\beta=\gamma=\delta=+$ and $\mu=\nu=-$. Relating the
inequalities (\ref{eqn:clusterbisepone}) and (\ref{eqn:clusterbiseptwo}) 
with each other as in the second 
case we obtain
\begin{eqnarray}
F_{+--+} \ge \sum_{ij, ij \not = --} F_{+ ij +} = F_{++}
&\mbox{ and }&
F_{- -- -} \ge \sum_{ij, ij \not = --} F_{- ij -} = F_{--}.
\end{eqnarray}
As a direct consequence we can split each of $F_{++++}$ and $F_{----}$ into two non negative parts
\begin{eqnarray}
F_{++++} = \tilde{F}_{++++} + F_{++},
&\mbox{ and }&
F_{----} = \tilde{F}_{----} + F_{--}.
\end{eqnarray}
With this definition of $\tilde{F}_{++++}$ and $\tilde{F}_{----}$
we can write
\begin{align}
\nonumber \varrho =& \tilde{F}_{++++} \ketbras{++++} + \tilde{F}_{----} \ketbras{----} 
\\
&+ \sum_{ij} F_{- ij +}\ketbras{- ij +} + F_{+ ij -} \ketbras{+ ij -} +\sigma_1,
\end{align}
where the state
\begin{eqnarray}
\sigma_1 =& \sum_{ij, ij \not = ++}F_{+ij+}(\ketbras{++++}+ \ketbras{+ ij +}) 
\nonumber
\\
& +\sum_{ij, ij \not = --}F_{- ij -}(\ketbras{----}+ \ketbras{- ij -})
\end{eqnarray}
is biseparable. With these replacements the largest violation 
of inequality (\ref{eqn:clusterbiseptwo}) is given by
\begin{equation}
\mathcal{V} \equiv \frac 1 2 \tilde{F}_{++++} + \frac 1 2 \tilde{F}_{----} - \frac 1 2 \sum_{ij} F_{+ ij -} + F_{- ij +}.
\end{equation}
One can decompose $\tilde{F}_{++++}$ and $\tilde{F}_{----}$ further into two non negative parts
\begin{eqnarray}
\tilde{F}_{++++} = F_{++++}^{\mathit{ent}} + F_{++++}^{\mathit{bs}}
&\mbox{ and }&
\tilde{F}_{----} = F_{----}^{\mathit{ent}} + F_{----}^{\mathit{bs}},
\end{eqnarray}
such that
\begin{eqnarray}
F_{++++}^{\mathit{bs}} +  F_{----}^{\mathit{bs}} &= \sum_{ij} F_{+ ij -} + F_{- ij +}
&\mbox{ and }
\frac{1}{2} F_{++++}^{\mathit{ent}} + \frac{1}{2} 
F_{----}^{\mathit{ent}}  = \mathcal{V}.
\end{eqnarray}
Using this decomposition of the coefficients one can write 
$\sigma_2 = F_{++++}^{\mathit{bs}}\ketbras{++++} +
F_{----}^{\mathit{bs}} + \sum_{ij} F_{+ ij -} \ketbras{+ ij -} +F_{- ij +} \ketbras{- ij +}$ as a convex combination of biseparable 
states of the form in Eq.~(\ref{eqn:clusterbs}). 
We can then split $\varrho$ into a genuinely multiparticle 
entangled part and a biseparable rest $\sigma_1+\sigma_2$
\begin{equation}
\varrho = F_{++++}^{\mathit{ent}} \ketbras{++++} + F_{----}^{\mathit{ent}} \ketbras{----} + \sigma_1 + \sigma_2.
\end{equation}
In this case this yields the last expected estimate on the GMN
\be
N_g(\varrho) \le 
F_{++++}^{\mathit{ent}} N_g(\ket{++++}) +  F_{----}^{\mathit{ent}} N_g(\ket{----}) = \mathcal{V}
\ee
So the maximal violation of the negative expectation values of the witnesses from Eqs.~(\ref{eqn:clusterwone}) and
(\ref{eqn:clusterwtwo}) gives upper bounds on the GMN, 
which proves the claim. As in the case of GHZ-diagonal states, 
the bound holds for both possible normalizations of the GMN.
\qed

\end{document}